\documentstyle [floats,preprint,epsf,aps,eqsecnum]{revtex}
\input psfig

\makeatletter
\def\footnotesize{\@setsize\footnotesize{10.0pt}\xpt\@xpt
\abovedisplayskip 10\p@ plus2\p@ minus5\p@
\belowdisplayskip \abovedisplayskip
\abovedisplayshortskip  \z@ plus3\p@
\belowdisplayshortskip  6\p@ plus3\p@ minus3\p@
\def\@listi{\leftmargin\leftmargini
\topsep 6\p@ plus2\p@ minus2\p@\parsep 3\p@ plus2\p@ minus\p@
\itemsep \parsep}}
\long\def\@makefntext#1{\parindent 5pt\hsize\columnwidth\parskip0pt\relax
\def\strut{\vrule width0pt height0pt depth1.75pt\relax}%
$\m@th^{\@thefnmark}$#1}
\long\def\@makecaption#1#2{%
\setbox\@testboxa\hbox{\outertabfalse %
\reset@font\footnotesize\rm#1\penalty10000\hskip.5em plus.2em\ignorespaces#2}%
\setbox\@testboxb\vbox{\hsize\@capwidth
\ifdim\wd\@testboxa<\hsize %
\hbox to\hsize{\hfil\box\@testboxa\hfil}%
\else %
\footnotesize
\parindent \ifpreprintsty 1.5em \else 1em \fi
\unhbox\@testboxa\par
\fi
}%
\box\@testboxb
} %
\global\firstfigfalse
\makeatother

\begin{document}

\preprint {UW/PT-95-10}

\title  {Quenched chiral perturbation theory for heavy-light mesons}

\author {Stephen R. Sharpe and Yan Zhang}

\address
    {%
    Department of Physics,
    University of Washington,
    Seattle, Washington 98195
    }%
\date {October 1995}

\maketitle
\vskip -20pt

\begin {abstract}%
{%
{%
\advance\leftskip  -2pt
\advance\rightskip -2pt

We formulate quenched chiral perturbation theory for heavy-light mesons
coupled to pions, and
calculate the one-loop chiral logarithmic corrections to $f_B$,
$f_{B_{s}}$, $B_B$ and $B_{B_{s}}$. We also calculate these corrections
for ``partially quenched'' theories.
In both theories, the chiral logarithms diverge in the chiral limit,
indicating that (partially) quenched theories should not be used to
study this limit.
Comparing the chiral logarithms to those in QCD,
we estimate the errors caused by (partial) quenching.
By forming suitable ratios, we can reduce the uncertainties in our estimates.

}%
}%
\end {abstract}
\newpage
\ifpreprintsty
\baselineskip=14pt
\fi
\section{Introduction}

Numerical simulations of lattice QCD are providing increasingly accurate
values for hadronic matrix elements.
Among the most interesting is the quantity $f_B^2 B_B$,
knowledge of which allows extraction of the CKM matrix element $V_{td}$
from experimental results for $B^0-\overline{B^0}$ mixing.
Unfortunately, present large scale simulations
use the ``quenched'' approximation, in which internal quark loops are dropped,
or the ``partially quenched'' approximation, in which the quarks in
the internal loops have a different mass from the valence quarks.
Unlike other sources of error, such as discretization,
it is difficult to systematically reduce the errors due to quenching,
Thus it is important to estimate these errors,
and in this paper we do so for the decay constants
and B-parameters of heavy-light mesons.

We estimate quenching errors by comparing the size of the
chiral logarithmic contributions to the quantities of interest in QCD
and quenched QCD (QQCD). Similarly, errors due to partial quenching
are estimated by comparing chiral logarithms in QCD and in partially
quenched QCD (PQQCD).
Chiral logarithms arise from the long-distance part of the
clouds of pseudo-Goldstone bosons (PGB's) surrounding all hadrons.
These clouds are radically different in the two theories,
there being, for example, a long-distance $\eta'$ cloud
around quenched hadrons, while there is no such cloud around hadrons in
QCD.\cite{Sharpelat89,Golterman1,Sharpe}
Thus, in general, there are significant differences in the chiral logarithms.
In particular, we find that $f_B$ and $B_B$ calculated in QQCD
should diverge in the limit that the light quark mass vanishes,
whereas in QCD they should approach a finite limit.
Such divergences are also predicted for light meson quantities,
e.g. $f_\pi$, and $m_\pi^2/m_q$ \cite{Golterman1,Sharpe},
and, indeed, have been observed \cite{Kim}.

Our estimates of quenching errors in individual quantities, such as $f_B$,
are necessarily crude, because chiral perturbation
theory tells us nothing about analytic contributions.
We can do much better, however, by considering ratios of quantities
designed so that all unknown contributions vanish.
The only corrections to these ratios come from $O(m_q^2)$
terms in chiral perturbation theory.


Our calculation is done by combining two technologies:
the chiral Lagrangian for heavy-light mesons \cite{Wise}
(which incorporates the heavy quark symmetry),
and chiral perturbation theory for the quenched and partially quenched
approximations \cite{Golterman1,BGPQ}.\footnote{%
For recent reviews of quenched chiral perturbation theory and its
applications, see Refs. \cite{Golterman2,Guptalat94}.}
Our results are thus valid up to corrections of $O(1/M_B)$,
which we expect to be small.

The outline of this paper is as follows. In the following section, we
explain how to extend quenched chiral perturbation theory
to include heavy-light mesons. In section
\ref{sec:calculation} we outline the calculation and give our results.
We discuss our estimates of the quenching errors in sec. \ref{sec:analysis},
and offer some conclusions in the final section.

After our calculations were completed,
a closely related paper by Booth appeared \cite{Booth1}.
He calculates the chiral logarithmic corrections to
$f_B$ and $B_B$ in QQCD, as well as
the corrections to $B$ masses and the Isgur-Wise function.
Our results for QQCD agree with his, except that we have
included two coupling constants that he overlooked.
This difference is not crucial, however, as these couplings are
expected to be small---indeed, in our numerical results we assume that
they vanish.
We keep the discussion of technical details to a minimum,
and stress results that are new.
In particular, we have (1) devised ratios in which unknown corrections
of $O(m_q)$ cancel, allowing a more reliable estimate of quenching errors
(this requires us to extend the calculations in QCD to the case of
non-degenerate quarks);
(2) considered wider ranges of the poorly known
parameters in the chiral Lagrangian; and
(3) extended the estimates to partially quenched QCD.

In a second work,
Booth has extended his calculations to include $1/M_B$ corrections
\cite{Booth2}.
This introduces a number of additional unknown constants, making
estimates of quenching errors more difficult.
It appears likely, however, that the $1/M_B$ corrections will not
be large enough to nullify the conclusions drawn from the infinite
mass limit assumed here.
A detailed discussion of these corrections in QCD has been given
by Boyd and Grinstein \cite{Boyd}.

\section{Formulation of quenched chiral lagrangian including
heavy-light mesons}
\label{sec:formulation}

We begin with a brief review of the quenched chiral Lagrangian
introduced in Ref. \cite{Golterman1}.
Following Morel \cite{Morel}, a functional integral formulation of QQCD
is obtained by adding, for each
light quark $q_{a}$ $(a=u,d,s)$, a ghost (bosonic) quark $\widetilde{q}_{a}$
with the same mass.
The fermion determinant of the quark sector
is then canceled by that of the ghost sector.
In the limit of vanishing quark masses,
and ignoring the anomaly for the moment,
the QQCD Lagrangian is
invariant under the graded group $U(3|3)_{L} \times U(3|3)_{R}$,
which mixes quarks and ghost-quarks.
Numerical evidence strongly suggests that this symmetry is broken
dynamically down to its vector subgroup, just as in QCD.
This gives rise to 36 Goldstone mesons, 9 each with compositions
$\phi_{ab} \sim q_a\overline{q}_b$,
$\widetilde{\phi}_{ab} \sim \widetilde{q}_a\overline{\widetilde{q}}_b$,
$\chi^{\dagger}_{ab} \sim q_a \overline{\widetilde{q}}_b$ and
$\chi_{ab} \sim \widetilde{q}_a \overline{q}_b$.
The first 18 are bosonic, the others fermionic.

The next step is to construct an effective Lagrangian describing the
low energy ($p\ll \Lambda_\chi\sim 1\,$GeV)
interactions of these particles.
We first collect the Goldstone fields into an hermitian $6 \times 6$ matrix,
\begin{equation}
\Phi=\left ( \begin{array}{cc}
              \phi & \chi^{\dagger} \\
              \chi & \widetilde{\phi}
             \end{array} \right ) \,,
\label{phi}
\end{equation}
which transforms linearly under the vector $U(3|3)$, but non-linearly
under the full chiral group.
Its exponential, however, transforms linearly
\begin{equation}
\Sigma=\exp(2i\Phi/f)\,, \quad
\Sigma \longrightarrow L \Sigma R^{\dagger} \,,
\label{sigma}
\end{equation}
where $L \in U(3|3)_{L}$ and $R \in U(3|3)_{R}$.
The lowest order
Lagrangian respecting the $U(3|3)_{L} \times U(3|3)_{R}$ symmetry is
\begin{equation}
 {\cal L}_{inv} = \frac{f^{2}}{8} {\sf str} (\partial_{\mu}\Sigma\partial^{\mu}
\Sigma^{\dagger}) + \lambda_{0}\, {\sf str} (M \Sigma+M \Sigma^{\dagger}),
\label{linv}
\end{equation}
where the mass term is
\begin{equation}
M={\rm diag}(m_u, m_d, m_s, m_u, m_d, m_s) \,,
\end{equation}
and the supertrace ${\sf str}$ appears because of the graded structure of
the symmetry group\cite{Dewitt}.
$f$ and $\lambda_{0}$ are undetermined bare constants.
In our numerical estimates, we set the scale in QQCD (and PQQCD) by taking
$f^{\rm QQCD}=f^{\rm PQQCD}=f^{\rm QCD}=130\,$MeV.
(This is slightly less than $f_\pi=132\,$MeV
because of the extrapolation to the chiral limit.)
Having made this choice,
we use the common symbol $f$ for all three quantities.
In the same spirit, since $\lambda_0$ does not appear
in the final expressions, we do not distinguish between its versions
in the three theories.

The next step is to account for the anomaly.
The chiral symmetry of the quantum theory is, in fact,
$[SU(3|3)_{L} \times SU(3|3)_{R}]
{\mathrel>\joinrel\mathrel\triangleleft}U(1)$.
In particular the field
\begin{equation}
\Phi_{0}=
{ (\eta^{\prime}-\widetilde{\eta^{\prime}})\over\sqrt{2}}
= { {\sf str} \Phi \over \sqrt6}
\propto
{\sf str} \ln \Sigma = \ln {\sf sdet} \Sigma
\end{equation}
($\eta^{\prime}$ is the normal $SU(3)$ singlet meson,
$\widetilde{\eta^{\prime}}$ its counterpart in the ghost sector)
is invariant under this smaller group, but not under the full
$U(3|3)_{L} \times U(3|3)_{R}$.
Thus we can multiply each term in the $O(p^2)$ quenched chiral Lagrangian
by an arbitrary function of $\Phi_0$. The full structure is
discussed in Ref. \cite{Golterman1}; in our one-loop calculations we need only
the terms quadratic in $\Phi_0$:
\begin{equation}
 {\cal L}_{light}={\cal L}_{inv} + \alpha (\partial_{\mu} \Phi_{0})^{2}
                  - m_0^2 \Phi_{0}^{2} \,.
\label{light}
\end{equation}
We are using the notation and conventions of Ref. \cite{Golterman1}
(in particular $f_\pi=132\,$MeV),
except that we use $m_0$ instead of $\mu$, following Ref. \cite{Sharpe2}.

It would appear that $\alpha$ and $m_0$ can be absorbed
into the quadratic part of ${\cal L}_{inv}$, renormalizing the fields
and masses of the $\eta'$ and $\widetilde\eta'$ particles.
In fact, as explained in Refs. \cite{Golterman1,Sharpe},
$\alpha$ and $m_0$ should be treated as interactions.
Diagrams with multiple insertions
of either coupling on an $\eta'$ or $\widetilde\eta'$ line cancel
(because they correspond in the underlying theory to
diagrams containing quark loops). Thus the $m_0$ term does not lead to
a mass for the $\eta'$ or $\widetilde\eta'$.
They remain light, and must be included
in the effective Lagrangian. This is in contrast to QCD, in which the
$\eta'$ gets a mass of $O(m_0)$ and can be integrated out.
In our calculation (as in all work in quenched
chiral perturbation theory to date) we assume that $\alpha$ and $m_0^2$
are small, so that we can work to one-loop order.

Diagonalizing the mass term in ${\cal L}_{inv}$, one finds that the
flavor diagonal mesons have compositions $u\overline u$,
$d\overline d$ and $s\overline s$, with masses
\begin{equation}
m_{uu}^{2}=\frac{8\lambda_{0}m_{u}}{f^{2}}\,,\quad
m_{dd}^{2}=\frac{8\lambda_{0}m_{d}}{f^{2}}\,,\quad
m_{ss}^{2}=\frac{8\lambda_{0}m_{s}}{f^{2}}\,,
\label{mss2}
\end{equation}
respectively. The flavor non-diagonal
particles have the same masses as in QCD, e.g.
$m_{K^+}^2 = 4\lambda_{0}(m_u + m_s)/f^2$.
The states containing ghosts have the same flavor compositions and masses
as those without.

This ends our review of the construction of Bernard and Golterman.
We now extend the effective quenched Lagrangian to include the
ground state mesons containing a heavy and a light quark.
As in the discussion above, the result is a simple extension of
that in QCD, which was worked out in Refs. ~\cite{Wise}.
The only differences are the presence of fields including ghost-quarks,
and extra interactions involving $\Phi_0$.

In the heavy quark symmetry limit the pseudoscalar and vector heavy-light
mesons are degenerate. They are destroyed, respectively, by the fields
$P_{a}^{(Q)}$ and $P_{a_{\mu}}^{*(Q)}$.
These fields have an implicit label $v$ denoting the four-velocity
of the corresponding particles.
In terms of the underlying quarks the flavor composition is
$P_a^{(Q)} \sim Q \overline{q}_a$, so that the meson destroyed by
$P_a^{(Q)}$ contains a heavy quark $Q$ and a light anti-quark of flavor $a$.
In QQCD this index runs over $u$, $d$, $s$ {\rm and}
$\widetilde{u}$, $\widetilde{d}$, $\widetilde{s}$.
Thus $P_{u,d,s}^{(Q)}$ are commuting fields, while
$P_{\widetilde{u},\widetilde{d},\widetilde{s}}^{(Q)}$ are anticommuting.
In the chiral limit all flavors are degenerate.

To display the heavy quark symmetry we form the spinor field
\begin{equation}
H_{a}^{(Q)}=\frac{1+\!\not\!v}{2}\left[P_{a_{\mu}}^{*(Q)} \gamma^{\mu}
           - P_{a}^{(Q)} \gamma_{5} \right ] \,.
\label{ha}
\end{equation}
which transforms as
\begin{equation}
H_{a}^{(Q)}\longrightarrow S H_{a}^{(Q)}\,,
\end{equation}
where $S$ is an element of the heavy quark spin $SU(2)_v$ group~\cite{Georgi}.
We do not elaborate on this symmetry here,
since the part of our construction which uses this symmetry is identical
to that for QCD.
To display the transformation properties under
$SU(3|3)_{L} \times SU(3|3)_{R}$ we follow the standard procedure
of introducing $U \in SU(3|3)$,
defined such that
\begin{equation}
\xi=\exp(i\Phi/f) \longrightarrow L\xi U^{\dagger}=U \xi R^{\dagger} \,,
\end{equation}
where now $L \in SU(3|3)_{L}$ and $R \in SU(3|3)_{R}$.
Then $H$ transforms as
\begin{equation}
H_{a}^{(Q)}\longrightarrow H_{b}^{(Q)}U_{ba}^{\dagger}\,.
\end{equation}
The conjugate field is defined by
\begin{equation}
\overline{H}_{a}^{(Q)}= \gamma^{0} {H_{a}^{(Q)}}^{\dagger}\gamma^{0} \,,
\qquad
\overline{H}_{a}^{(Q)} \longrightarrow
(U_{ab} \overline{H}_{b}^{(Q)}) S^{-1} \,.
\end{equation}
This field creates heavy-light mesons of
velocity $v$, and has nothing to do with the anti-particles of these mesons.

Now we can write down the part of the
effective Lagrangian containing $H_a^{(Q)}$
\begin{eqnarray}
{\cal L}_{Q} &=&-i \, {\sf str}_{a}Tr\overline{H}_{a}^{(Q)}
	v\cdot \partial H_{a}^{(Q)}
        + {\textstyle{1\over2}}
	i {\sf str}_{a} Tr \overline{H}_{a}^{(Q)} H_{b}^{(Q)} v^{\mu}
        \left[ \xi^{\dagger} \partial_{\mu} \xi + \xi \partial_{\mu}
        \xi^{\dagger} \right]_{ba} \nonumber \\
       & & + {\textstyle{1\over2}}
	i \, g \, {\sf str}_{a} Tr \overline{H}_
        {a}^{(Q)} H_{b}^{(Q)} \gamma_{\mu} \gamma_{5} \left[
        \xi^{\dagger} \partial^{\mu} \xi - \xi \partial^{\mu}\xi^{\dagger}
        \right]_{ba} \nonumber \\
       & & + {\textstyle{1\over2}}
	i \, g\prime \, {\sf str}_{a}
        Tr \overline{H}_{a}^{(Q)} H_{a}^{(Q)} \gamma_{\mu} \gamma_{5} \,
        {\sf str} \left[ \xi^{\dagger} \partial^{\mu} \xi -
        \xi \partial^{\mu}\xi^{\dagger} \right] + \cdots \,,
\label{lq}
\end{eqnarray}
where the ellipsis denotes terms of higher order in momentum, quark masses,
and $1/m_{Q}$.
${\sf str}_{a}$ indicates that the supertrace is to be taken
over the repeated index $a$.
Summation over other repeated indices is assumed.
Finally, the trace $Tr$ is taken over the Dirac indices.
${\cal L}_{Q}$ is invariant under the chiral group, the heavy quark
spin symmetry, and, of course, Lorentz transformations.
In principle, each term in ${\cal L}_{Q}$ can be multiplied
by a real and even but otherwise arbitrary function of $\Phi_{0}$.\footnote{%
Note that a possible term linear in
$\Phi_0$ of the form $\Phi_0 {\sf str}_a Tr \overline{H}_a H_a \gamma_5$
vanishes identically.}
We do not display these functions as they contribute only at two-loop order.
Aside from these (implicit) functions, the only difference between the
quenched Lagrangian ${\cal L}_Q$ and that in QCD is the appearance of
the $g\prime$ term.
Since ${\sf str} \left[ \xi^{\dagger} \partial^{\mu} \xi -
        \xi \partial^{\mu}\xi^{\dagger} \right] \propto \partial^\mu\Phi_0$,
this term gives rise to a coupling of
the $\eta'$ and $\widetilde\eta'$ to the heavy-light mesons.
Though such a coupling exists in QCD, it is not needed in the QCD chiral
Lagrangian because the $\eta'$ is integrated out.

To discuss $B$-parameters,
we also need the Lagrangian describing the interactions of PGB's with
the antiparticles of the heavy-light mesons, i.e. those with composition
$q_a\overline{Q}$. We follow Ref.~\cite{Grinstein}
and introduce a field which destroys such mesons
\begin{equation}
H_{a}^{(\overline{Q})}=\left[P_{a_{\mu}}^{*(\overline{Q})} \gamma^{\mu}
           - P_{a}^{(\overline{Q})} \gamma_{5} \right]\frac{1-\!\not\!v}{2} \,.
\end{equation}
This transforms under heavy quark spin symmetry and
$SU(3|3)_{L} \times SU(3|3)_{R}$ as
\begin{equation}
H_{a}^{(\overline{Q})}\longrightarrow (U_{ab}H_{b}^{(\overline{Q})})S^{-1}\,.
\end{equation}
The conjugate field $\overline{H}^{(\overline{Q})}$ is defined as for
$H^{(Q)}$. The invariant Lagrangian is
\begin{eqnarray}
{\cal L}_{\overline{Q}} &=& -i \, {\sf str}_{a}Tr \,
	v\cdot \partial H_{a}^{(\overline{Q})}
        \overline{H}_{a}^{(\overline{Q})} -
	{\textstyle{1\over2}}
	i {\sf str}_{a} Tr H_{a}^{(\overline{Q})}
        \overline{H}_{b}^{(\overline{Q})}v^{\mu} \left[
        \xi^{\dagger} \partial^{\mu} \xi + \xi \partial^{\mu}\xi^{\dagger}
	\right]_{ba}
        \nonumber \\
        & & + {\textstyle{1\over2}}
	i \, g \, {\sf str}_{a} Tr
        H_{a}^{(\overline{Q})} \overline{H}_{b}^{(\overline{Q})}
	\gamma_{\mu} \gamma_{5}
	\left[
        \xi^{\dagger} \partial^{\mu} \xi - \xi \partial^{\mu}\xi^{\dagger}
        \right]_{ba} \nonumber \\
	& & + {\textstyle{1\over2}}
	i\, g\prime \, {\sf str}_{a}
        Tr H_{a}^{(\overline{Q})} \overline{H}_ {a}^{(\overline{Q})}
	\gamma_{\mu} \gamma_{5}
        {\sf str} \left[ \xi^{\dagger} \partial^{\mu} \xi -
        \xi \partial^{\mu}\xi^{\dagger} \right] + \cdots \,.
\end{eqnarray}
The total quenched chiral Lagrangian is then
${\cal L}={\cal L}_{light} + {\cal L}_{Q} + {\cal L}_{\overline{Q}}$.

We are interested in decay constants and $B$-parameters.
These are defined, respectively, by
\begin{eqnarray}
\langle 0| \overline{u}\gamma_{\mu}(1\!-\!\gamma_{5})b|B^{-}(v) \rangle
& = & -if_{B}p_{\mu} = -i f_B m_B v_\mu \,,
\label{dfb} \\
\langle \overline{B^{0}}(v)|
	\overline{b}\gamma^{\mu}(1\!-\!\gamma_{5})d\
	\overline{b}\gamma_{\mu}(1\!-\!\gamma_{5})d      |B^{0}(v) \rangle
&=& \frac{8}{3}f^{2}_{B}m^{2}_{B}B_{B}\,,
\label{eq:BB}
\end{eqnarray}
with similar equations for mesons of other flavors.
We thus need to know how the left-handed currents,
$L_{a}^{\mu}=\overline{q}_{a} \gamma^{\mu}(1\!-\!\gamma_{5})Q$,
and the four-fermion
operators composed of two such currents, can be written in terms of
the fields in the quenched chiral Lagrangian. The result
is that the matrix elements of the current involving the
$Q$ field are given by
\begin{equation}
L_{a}^{\mu}=\frac{i\kappa}{2} Tr \gamma^{\mu}(1-\gamma_{5})H_{b}^{(Q)}
            \xi_{ba}^{\dagger} V_L(\Phi_0) + \cdots
\label{lhcurrent}
\end{equation}
where the ellipsis denotes non-leading operators in the chiral and heavy
quark expansions.
The ``potential'' $V_L$ can contain both odd and even powers of $\Phi_0$,
with arbitrary coefficients---and we find that the linear term
does contribute at one-loop order.
We normalize the potential so that $V_L(0)=1$.\footnote{%
We note in passing that similar potentials occur in the chiral representation
of currents composed of light quarks.
They do not, however, give rise to one loop corrections
to $f_\pi$ and $f_K$.
This is because CP, together with quark-interchange symmetries,
require that the potentials be even functions of $\Phi_0$.}
Then, taking into account the factor of
$\sqrt{m_{Q\overline{q}}}$ absorbed into the fields $P^{(Q)}$,
we have $\kappa=f_{Q\overline{q}} \sqrt{m_{Q\overline{q}}}$ at leading order.
All this is just as in QCD\cite{Wise},
except that the index $a$ now runs from $1-6$.

Similarly, following Ref. \cite{Grinstein},
the mixing matrix elements of the $\Delta Q=2$ operator
$O_{aa}=\overline{Q}\gamma_{\mu}(1- \gamma_{5})q_{a}\overline{Q}
\gamma^{\mu}(1- \gamma_{5})q_{a}$ (no sum on $a$) can be represented
in the effective theory by the operator
\begin{equation}
O_{aa}=\sum_{b,c}
\beta Tr\xi_{ab}\overline{H}_{b}^{(Q)} \gamma_{\mu}(1-\gamma_{5}) \,
       Tr\xi_{ac}H_{c}^{(\overline{Q})}\gamma^{\mu}(1-\gamma_{5}) V_O(\Phi_0)
 + \cdots
\end{equation}
Again, the arbitrary potential is normalized so that $V_O(0)=1$,
and can contain terms linear in $\Phi_0$ which will contribute at one-loop
order.
This representation is not unique, but, as explained in Ref. \cite{Grinstein},
other choices are equivalent since they all lead,
after evaluating the Dirac traces, to the form
\begin{equation}
O_{aa}=4\beta \left[ \left(\xi P_{\mu}^{*(Q)\dagger}\right)_{a}
         \left(\xi P^{*(\overline{Q})\mu}\right)_{a} + \left(\xi P^{(Q)\dagger}
         \right)_{a}\left(\xi P^{(\overline{Q})}\right)_{a}\right] V_O(\Phi_0)
+ \cdots
\end{equation}
At leading order $\beta=\frac23 f_B^2 m_B B_B$.

\subsection{Partially quenched theories}
\label{subsec:pq}

Partially quenched theories are a first step towards removing the quenched
approximation. In these theories, not all the quarks are quenched, so that
there are more quarks than ghost quarks. The particular example we focus
on is that used in recent simulations (e.g. see Ref. \cite{bernardlat95}):
the lattices are generated with $N_f$ quarks of mass $m_f$, while
valence quarks of mass $m_q$ are used to calculate $f_{B_q}$ and $B_{B_q}$.
Although $N_f=2$ in present simulations, we keep $N_f$ as a free parameter,
so as to make our results more general.
In the limit that $m_q=m_f$ the theory is unquenched and QCD-like,
having $N_f$ flavors of degenerate quarks.
Thus we refer to it as partially quenched QCD.

It is straightforward to generalize the development of the
previous subsection to PQQCD.
Most of the work has been done by Bernard and Golterman,
who developed chiral perturbation theory for partially quenched
theories \cite{BGPQ}.
There are only two alterations.
\begin{itemize}
\item
The index $a$ labeling quarks and ghost-quarks now takes on
$N_f+2$ values: $a=(q, f_1, \dots, f_{N_f}, \widetilde{q})$.
(We could introduce quenched $u$, $d$ and $s$ quarks, as in our
discussion of QQCD, but this does not affect the results for the
quantities of interest.)
The initial symmetry is thus $U(N_f+1|1)_L \times U(N_f+1|1)_R$,
the fields $\Phi$, $\Sigma$ and $\xi$ being $(N_f+2)\times (N_f+2)$ matrices.
The form of the heavy-light meson Lagrangian is unchanged.
\item
The $\alpha$ and $m_0^2$ vertices can be iterated in
the $SU(N_f)$ sector, so that the mass of the $\eta'$ in this sector
is shifted from the quark mass contribution,
$m_{ff}^2= {8 \lambda_0 m_f / f^2}$, to
\begin{equation}
m_{\eta'}^2 = {m_{ff}^2 + m_0^2 {N_f / 3} \over
		1 + \alpha {N_f / 3} } \,.
\end{equation}
The $q\overline{q}$ meson mixes with this $\eta'$ through the $m_0^2$ and
$\alpha$ vertices, leading to a propagator with a complicated pole
structure, given explicitly in Ref. \cite{BGPQ}.
\end{itemize}
To simplify the resulting expressions, we assume that both $m_f$ and $m_q$
are small,
and that $N_f$ is large enough that the $SU(N_f)$ $\eta'$ is heavy.
We then have a hierarchy of scales
\begin{equation}
m_{qq}^2 = {8 \lambda_0 m_q \over f^2} \sim m_{ff}^2 \ll
m_0^2 {N_f \over 3} \sim \Lambda \,,
\end{equation}
and we can drop terms suppressed by $m_{qq}^2/(m_0^2 N_f/3)$
or $m_{ff}^2/(m_0^2 N_f/3)$.
When this is done, all contributions that depend on the
``quenched couplings'' (i.e. $m_0^2$, $\alpha$, $g\prime$,
$V_L'(0)$ and $V_O'(0)$) turn out to be higher order in
the chiral expansion than those we keep.
It should be borne in mind, however, that, for the largest values of
$m_{ff}^2$ used in present simulations,
corrections proportional to $m_{ff}^2/m_0^2$ may be substantial.

\section{Calculation of Chiral Logarithms}
\label{sec:calculation}

\begin{figure}
\begin{center}
\begin{picture}(430,230)
\put(65,116){(a)}
\put(210,116){(b)}
\put(355,116){(c)}
\put(65,0){(d)}
\put(210,0){(e)}
\put(355,0){(f)}
\linethickness{1mm}
\multiput(15,145)(145,0){3}{\line(1,0){108}}
\multiput(15,29)(145,0){3}{\line(1,0){108}}
\thinlines
\linethickness{8pt}
\multiput(122,145)(145,0){3}{\line(1,0){8}}
\multiput(122,29)(145,0){3}{\line(1,0){8}}
\thinlines
\multiput(99,145)(145,0){2}{\oval(56,56)[t]}
\put(359,145){\oval(54,54)[t]}
\put(69,29){\oval(54,54)[t]}
\multiput(271,44)(145,0){2}{\circle{30}}
\thicklines
\put(239,168){\line(1,1){10}}
\put(239,178){\line(1,-1){10}}
\put(64,51){\line(1,1){10}}
\put(64,61){\line(1,-1){10}}
\put(411,54){\line(1,1){10}}
\put(411,64){\line(1,-1){10}}
\thinlines
\multiput(71,145)(145,0){2}{\circle*{8}}
\multiput(332,145)(54,0){2}{\circle*{8}}
\multiput(42,29)(54,0){2}{\circle*{8}}
\end{picture}
\end{center}
\caption{ \label{fig1}
	Graphs contributing to the one-loop renormalization of the decay
        constant. Thick and thin lines are propagators for heavy-light mesons
	and PGB's respectively. Solid circles are vertices proportional
 	to $g$ or $g\prime$, crosses are insertions of the interactions
	proportional to $\alpha$ or $m_0$, and the filled squares denote
	the left-handed current.
        }
\end{figure}
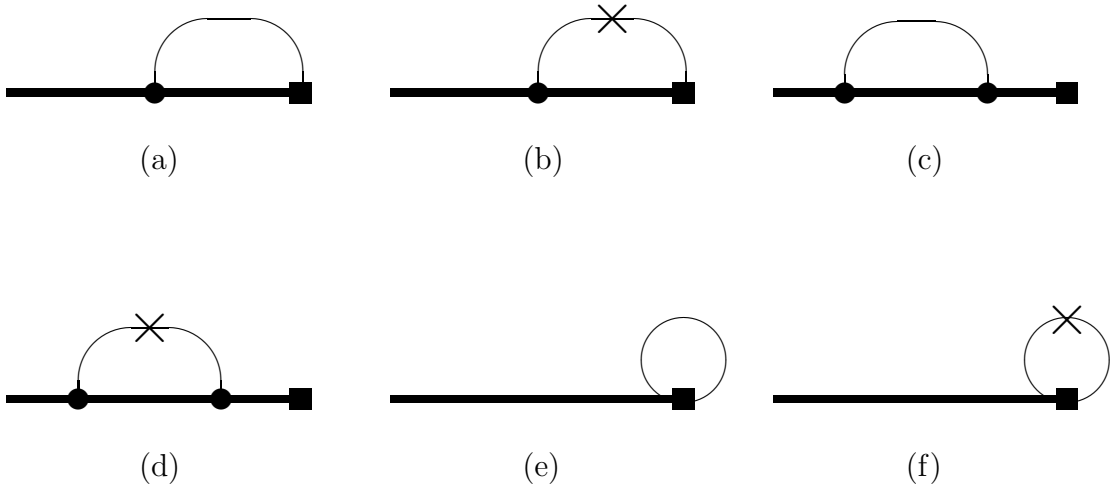

We first calculate the one-loop chiral logarithms contributing to
$f_{Q \overline{s}}$.
Strictly speaking, we calculate the logarithms for the quantity
$\kappa_s \sqrt{m_{Q\overline{s}}}= f_{Q \overline{s}} {m_{Q \overline{s}}}$.
At leading order in $1/m_Q$, however,
these results apply also for the decay constants,
because chiral logarithms in $m_{Q\overline{s}}$ are suppressed by $1/m_Q$.
The graphs which contribute are shown in Fig.~(\ref{fig1}).
The calculations are standard---the details are as in QCD \cite{Grinstein}.
In particular, the heavy-quark symmetry allows us to choose the external meson
to be either the $P$ or $P^*$. We choose the former, which allows
us to extract the result by contracting the external current with $v_\mu$.
The three-point vertices in ${\cal L}_Q$ connect a $P$ only to a $P^*$,
so that the heavy-light meson in the loops is always the $P^*$.
This means that graphs (a) and (b) vanish identically (as in QCD),
because the factor of $v_\mu$ from the current annihilates the
$P^*$ propagator.

The absence of quark loops in QQCD manifests itself as cancelations between
intermediate states containing ghosts and those without.
Thus, while graph (c) can give rise to contributions proportional
to $g^2$, $g\prime^2$, and $g g\prime$, only the latter is non-vanishing.
Similarly, the only non-canceling contribution from (d) is that
proportional to $g^2$, and the contributions from (e) cancel entirely.
The net effect is that only loops involving the $\overline{s}s$ PGB
contribute---just as expected since the $s$ quark is the only light quark
available. Expressing the decay constant as
\begin{equation}
\left(f_{Q\overline{s}}\right)^{{\rm QQCD}} =
{\kappa^{\rm QQCD} \over \sqrt{m_{Q\overline{s}}}} \,
\left[1 + (\Delta f_{Q\overline{s}})^{{\rm QQCD}}
+ c_1^Q(\Lambda) m_0^2 + c_2^Q(\Lambda) m_s + \dots
\right] \,,
\label{eq:fs}
\end{equation}
we find the chiral logarithm to be
\begin{equation}
(\Delta f_{Q\overline{s}})^{{\rm QQCD}} =
      \frac{1}{(4\pi f)^{2}} \left[\frac{1}{6}(1+3g_Q^{2})
      \left(2\alpha m_{ss}^{2} -m_0^{2} \right)
	- \left( {i f V_L'(0) \over \sqrt6} + 3 g\prime g_Q\right)
      m_{ss}^{2} \right] \log\frac{m_{ss}^{2}}{\Lambda^{2}}  \,.
\label{eq:Dfds}
\end{equation}
Here $m_{ss}^{2}$ is defined in Eq.~(\ref{mss2}),
and $\Lambda$ is the ultraviolet cut-off on the logarithmic divergences.
The $\Lambda$ dependence in $\Delta f$
is canceled by that of the analytic terms
proportional to $c_{1,2}^Q$ in Eq. \ref{eq:fs}.
These constants result from higher order terms in the expression for
the current $L^\mu$ and in the heavy-light meson Lagrangian.
Their structure is discussed by Booth \cite{Booth1}.

The quenched result for $f_{Q\overline{q}}$ ($q=u,d$) is obtained from
Eq. \ref{eq:Dfds} simply by substituting $m_{qq}$ for $m_{ss}$.
Present lattice simulations, however,
do not directly work with physical $u$ and $d$ quarks,
but instead extrapolate from heavier quark masses.
We mimic this procedure analytically in the next section.

The entire result in Eq. \ref{eq:Dfds} is a quenched artifact, being
proportional to one of $\alpha$, $m_0^2$, $g\prime$ or $V_L'(0)$.
The quenched logarithms are thus very different from those in QCD.
In particular, the term proportional to $m_0^2$ diverges in the chiral limit,
whereas the chiral logarithms in QCD,
to which we now turn, vanish in this limit.

The chiral logarithms have been calculated in QCD
in the limit $m_u=m_d=0$ in Ref. \cite{Grinstein}.
We need, however,
the result for the general case of three non-degenerate quarks.
We use a similar expansion to that for QQCD
\begin{equation}
\left(f_{Q\overline{q}}\right)^{{\rm QCD}} =
{\kappa^{\rm QCD} \over \sqrt{m_{Q\overline{q}}}} \,
\left[1 + (\Delta f_{Q\overline{q}})^{{\rm QCD}}
+ c_1(\Lambda) (m_u+m_d+m_s) + c_2(\Lambda) m_q + \dots
\right] \,,
\label{eq:fullfs}
\end{equation}
where $q=u$, $d$ or $s$.
Note that the analytic terms have a somewhat different form to those in QQCD.
The chiral logarithm comes from Figs. \ref{fig1} (c) and (e), and is
\begin{equation}
\label{eq:fullfqlog}
(\Delta f_{Q\overline{q}})^{{\rm QCD}} =
	- \frac{1}{(4\pi f)^{2}} (1+3g^{2})
	\sum_{a} (T_a^2)_{qq} m_a^2 \log\frac{m_a^2}{\Lambda^2} \,.
\end{equation}
The sum runs over the PGB's (eight in QCD) in the mass eigenbasis,
with $T_a$ being the corresponding group generator,
normalized such that
${\sf tr}[T_a T_b] = {\textstyle{1\over2}} \delta_{ab}$.
This formula hides the fact that, for $m_u\ne m_d$,
the $\pi^0$ and $\eta$ mix, so that one must diagonalize the mixing
matrix to determine the appropriate $T_a$'s to use.

The result for PQQCD is obtained from the same graphs as that in QQCD,
except that the crosses in Fig. \ref{fig1}(b) and (d) now include
a projection against the $SU(N_f)$ $\eta'$.
The expansion for the decay constant is
\begin{equation}
\left(f_{Q\overline{q}}\right)^{{\rm PQQCD}} =
{\kappa^{\rm PQQCD} \over \sqrt{m_{Q\overline{q}}}} \,
\left[1 + (\Delta f_{Q\overline{q}})^{{\rm PQQCD}}
+ c_1^{PQ}(\Lambda) m_f + c_2^{PQ}(\Lambda) m_q + \dots
\right] \,,
\label{eq:pqfs}
\end{equation}
with the chiral logarithm being
\begin{equation}
\label{eq:pqfqlog}
(\Delta f_{Q\overline{q}})^{{\rm PQQCD}} =
	- \frac{1}{(4\pi f)^{2}} (1+3g_{PQ}^{2}) \left(
{N_f\over 2} m_{qf}^2 \log\frac{m_{qf}^2}{\Lambda^2}
+ {1 \over 2N_f} (m_{ff}^2 - 2 m_{qq}^2) \log\frac{m_{qq}^2}{\Lambda^2}
\right) \,.
\end{equation}
Here $m_{qf}$ is the mass of the meson composed of a valence quark
(mass $m_q$) and a dynamical antiquark (mass $m_f$), and to the order
we work is given by $m_{qf}^2 = (m_{qq}^2+m_{ff}^2)/2$.
A check of Eq. \ref{eq:pqfqlog} is that,
in the limit $m_q=m_f$, it equals that for QCD with $N_f$ degenerate flavors
(easily obtained from Eq. \ref{eq:fullfqlog}).
We stress again that terms proportional to $\alpha$, $m_0^2$ and $g\prime$
are absent.

\begin{figure}
\begin{center}
\begin{picture}(285,230)
\put(65,116){(a)}
\put(210,116){(b)}
\put(65,0){(c)}
\put(210,0){(d)}
\linethickness{8pt}
\multiput(66,145)(145,0){2}{\line(1,0){8}}
\multiput(66,29)(145,0){2}{\line(1,0){8}}
\thinlines
\linethickness{1mm}
\multiput(15,145)(145,0){2}{\line(110,0){110}}
\multiput(15,29)(145,0){2}{\line(110,0){110}}
\thinlines
\multiput(70,145)(145,0){2}{\oval(56,56)[t]}
\multiput(70,44)(145,0){2}{\circle{30}}
\multiput(42,145)(145,0){2}{\circle*{8}}
\multiput(98,145)(145,0){2}{\circle*{8}}
\thicklines
\put(210,168){\line(1,1){10}}
\put(210,178){\line(1,-1){10}}
\put(210,54){\line(1,1){10}}
\put(210,64){\line(1,-1){10}}
\thinlines
\end{picture}
\end{center}
\caption{ \label{fig3}
           Graphs contributing to the one-loop renormalization of
           $B_{Q\overline{q}}$. The solid square denotes the operator
           $O_{aa}$.  Other notation as in Fig. 1.
         }
\end{figure}
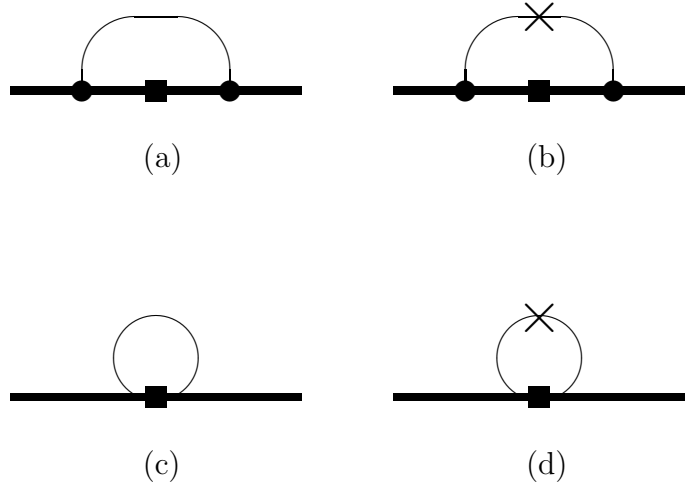

We now turn to results for $B$-parameters,
the relevant graphs for which are shown in Fig. \ref{fig3}.
We define $\Delta B$ in the same way as $\Delta f$
(Eqs. \ref{eq:fs}, \ref{eq:fullfs} and \ref{eq:pqfs})---one
can show that the analytic terms have the same form as for decay
constants for all three theories.
The values of the constants $c^Q$ for $B$-parameters are, of course,
not related to those for decay constants.
Our results for the chiral logarithms are
\begin{eqnarray}
(\Delta B_{Q\overline{q}})^{{\rm QQCD}} & = &
      -   \frac{1}{(4\pi f)^{2}} \left[(1-3g_Q^{2})\left(m_{qq}^{2}
      -\frac{2\alpha}{3}m_{qq}^{2} +\frac{m_0^{2}}{3} \right) \right.
\nonumber \\
&&\mbox{}-\left(
	\left.{i 2 f\over \sqrt{6}} [V_O'(0)+V_L'(0)] + 6 g\prime g_Q\right)
	m_{qq}^{2} \right]\log\frac{m_{qq}^{2}}{\Lambda^{2}}
	\,,
\label{eq:Dbbs} \\
(\Delta B_{Q\overline{q}})^{{\rm QCD}} &=&
      - \frac{1}{(4\pi f)^{2}} (1-3g^{2})
	\sum_a 2 \left[(T_a)_{qq}\right]^2
		m_a^2 \log\frac{m_a^2}{\Lambda^2} \,,
\label{eq:fullbbq} \\
(\Delta B_{Q\overline{q}})^{{\rm PQQCD}} &=&
      - \frac{1}{(4\pi f)^{2}} (1-3g_{PQ}^{2}) \left[
	 m_{qq}^2 + {1\over N_f} (m_{ff}^2 - 2 m_{qq}^2) \right]
	\log\frac{m_{qq}^2}{\Lambda^2} \,.
\label{eq:pqbbq}
\end{eqnarray}
Only flavor-neutral PGB's contribute in all three theories.
This is explicit for QQCD and PQQCD, while for QCD it follows
because the expression picks out generators whose $qq$-th element
is non-vanishing.
As required, the PQQCD result reduces to that for QCD with $N_f$ degenerate
quarks when $m_f=m_q$.

The quenched result for $\Delta B$ is similar to that for $\Delta f$
(Eq. \ref{eq:Dfds}), except that the first term is not a quenched
artifact. It is of the form $m_{qq}^2 \log m_{qq}$,
but is not proportional to $\alpha$.
Indeed, this term is exactly the contribution that would be obtained from
the QCD formula, Eq. \ref{eq:fullbbq},
if one were to use the quenched spectrum of flavor-neutral mesons,
i.e. states with composition $u\overline{u}$,
$d\overline{d}$, and $s\overline{s}$.
In other words, the quenched result consists of that from QCD,
modified because the spectrum of PGB's is different,
together with contributions which are
quenched artifacts and involve the $\eta'$.
This is the same situation as for $B_K$, and can be understood
by an analogous argument using quark-line diagrams \cite{Sharpe}.

\section{Analysis and Discussion}
\label{sec:analysis}

To use our results we need to know the numerical values of the
dimensionless parameters $m_0/f$, $\alpha$, $g_Q$, $g_{PQ}$, $g\prime$,
$f V_L'(0)$ and $f V_O'(0)$.
These are parameters of the (partially) quenched theory, and can,
in principle, be determined by simulations.
At present, however, these parameters are not well known,
so we allow them to vary in ``reasonable'' ranges,
determined as explained in the following.
To compare to QCD we also need to know $g$.

There is now evidence for a non-zero value of $m_0/f$
\cite{Kuramashi,Kim,Guptalat94}. The result is conventionally expressed
in terms of $\delta=2 m_0^{2}/({3 (4 \pi f)^{2}})$.
A reasonable range for $\delta$ appears to be $0.05-0.15$.

The evidence for $\alpha$ is less clear. Like $m_0^2$, it is
proportional to $1/N_c$, and thus, hopefully, small.
The simplest choice is to assume $\alpha=0$ \cite{Golterman1,Sharpe}.
A better estimate, however, can be obtained from the data of
Ref.\cite{Kuramashi} on the $\eta'$ two-point function.
Fitting the data to the form expected from ${\cal L}_{light}$
(Eq. \ref{light}),
one finds $\alpha \simeq 0.7$ and $m_0 \simeq 720 MeV$ ($\delta=0.13$).
This result may have large discretization errors since it is obtained
at a relatively large lattice spacing ($\beta=5.7$).
Nevertheless, it suggests a positive value for $\alpha$.
We assume that $\alpha$ lies in the range $0-0.7$.

There are no lattice results concerning $g_Q$ or $g_{PQ}$,
so we assume that the value is similar to $g$.
Experimental results for $\Gamma(D^{*}\rightarrow D \pi)$ imply
$0.1<g^{2}< 0.5$ (see the discussion in Ref. \cite{Booth1}).
We use this as our reasonable range, and take $g_Q^2=g_{PQ}^2=g^2$.
To be consistent with Ref. \cite{Wise} we use
$g^{2}=0.4$ as our preferred value.

Finally, we come to $g\prime$, $V_L'(0)$ and $V_O'(0)$,
about which even less is known.
$g\prime$ is the additional $P^{(Q)}$-$P^{*(Q)}_{\mu}$-$\eta\prime$
coupling arising from the fact that the $\overline{q}q$ pair in
the $\eta\prime$ can annihilate into gluons.
It follows that $g\prime/g_Q \propto 1/N_c$,
though we have no information concerning the relative sign of $g_Q$ and
$g\prime$. We consider two possibilities:
a conservative choice, $-g_Q \leq g\prime \leq g_Q$,
and the more likely situation in which $g\prime$ is small,
$|g\prime/g_Q| \ll 1$.
Similarly, $V_L'(0)$ and $V_O'(0)$ correspond to additional couplings
of the $\eta'$ to the current and four-fermion operators, respectively,
involving annihilation into gluons. Thus both are of $O(1/N_c)$.
In the following, we assume that both vanish.
In fact, $g\prime$ and $V_L'(0)$ appear in the same way in both
the correction for the decay constant and $B$-parameter,
so that $V_L'(0)$ can be absorbed into $g\prime$.

\begin{figure}[tb]
\centerline{\psfig{file=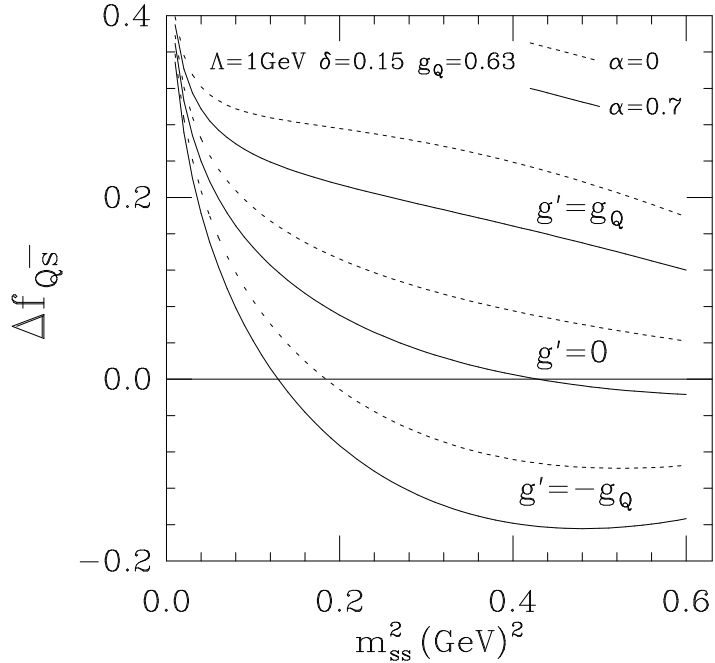,height=4.5truein}}
\vspace{-0.5truein}
\caption{Quenched chiral logarithmic contribution to $f_{Q\overline{s}}$.}
\label{fig:DfB2}
\end{figure}

With these parameters in hand,
we begin by plotting $(\Delta f_{Q\overline{s}})^{{\rm QQCD}}$
(Eq. \ref{eq:Dfds}),
as a function of $m_{ss}^2$. Here ``$s$'' indicates a generic
quenched ``light'' quark.
For the couplings we take $\delta=0.15$, $g_Q^2=0.4$,
$\Lambda=1\,$GeV,
and display the results for $\alpha=0,0.7$ and $g\prime=-g_Q,0,g_Q$.
The following important features turn out to hold for all
values of the couplings in our reasonable ranges.
\begin{itemize}
\item
The correction diverges as $m_{ss}\to 0$. This
reflects the unphysical cloud of light $\eta'$ particles
surrounding the quenched heavy-light mesons.
Observing this divergence in numerical results would give further
confirmation of the validity of quenched chiral perturbation theory.
In practice, this may be difficult, since the divergence does not
set in until $m_{ss}$ falls below $\sim 100\,$MeV.

\item
For masses heavy enough to be away from the divergence,
the correction can be substantial.
This is the region in which most present simulations are carried out:
``light'' quarks range in
mass from roughly $m_s^{\rm phys}/3$ to $m_s^{\rm phys}$,
so that $0.17 ({\rm GeV})^2 \le m_{ss}^2 \le 0.5 ({\rm GeV})^2$.
For these masses, and for couplings in our reasonable ranges,
we find that the magnitude of the correction can be as large as $30\%$.
If $|g\prime/g| \ll 1$, however,
the correction is smaller, not exceeding $15\%$.

\item
The departures from linearity in the range
$0.17 ({\rm GeV})^2 \le m_{ss}^2 \le 0.5 ({\rm GeV})^2$ are
fairly small, particularly for $g\prime\ge 0$.
In the following, we mimic numerical simulations, and define a
quenched ``$f_B$'' by linear extrapolation from the masses
$m_{ss}^2=0.25\,({\rm GeV})^2$ and $0.5\,({\rm GeV})^2$.

\item
$\Delta f_{Q\overline{s}}$ decreases with increasing $m_{ss}$.
Using the definition of $f_B$ just given,
the chiral logarithm alone thus implies $f_{B_s}/f_B<1$ in QQCD.
This is in contrast to QCD, where, for $g=0.63$ and $\Lambda=1\,$GeV,
$(\Delta f_{B_s})^{\rm QCD}=0.38>(\Delta f_B)^{\rm QCD}=0.21$.
We return to this point below.
\end{itemize}

\begin{figure}[tb]
\centerline{\psfig{file=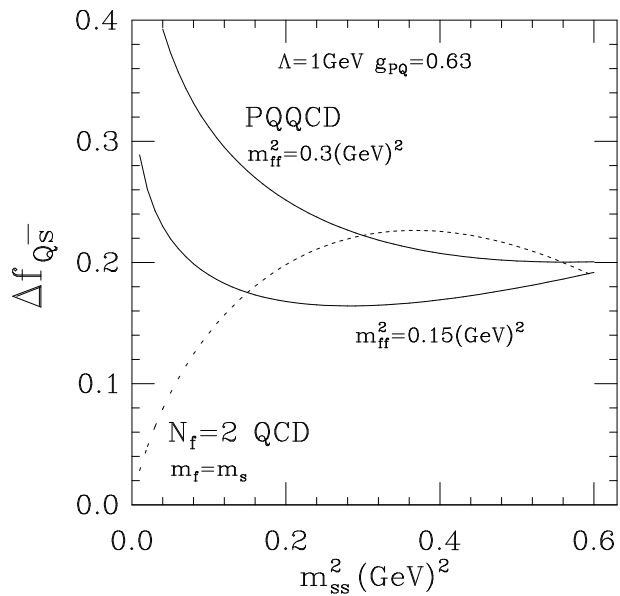,height=4.truein}}
\vspace{-0.5truein}
\caption{Chiral logarithms for $N_f=2$ dynamical quarks:
PQQCD with $m_f$ fixed at two values, and the QCD-like theory with $m_f=m_s$.
Parameters are $\Lambda=1\,$GeV and $g_{PQ}=0.63$.}
\label{fig:DBBpq}
\end{figure}

In Fig. \ref{fig:DBBpq} we show the size of the chiral logarithms
in partially quenched QCD.
We have chosen two values for the dynamical
pion mass which are in the range used in present simulations,
$m_{ff}^2=0.15$ and $0.3 {\rm GeV}^2$.
The divergence at small $m_{ss}$ remains,
although now it is proportional to $m_{ff}^2$ rather than $m_0^2$.
Again the corrections are substantial, although the dependence on
$m_{ss}^2$ is weaker than in QQCD.
We also show results for the unquenched
two flavor QCD-like theory obtained by setting $m_f=m_s$.
Although there is no divergence at small $m_{ss}$, the curve is not
very different from those for PQQCD in the mass range of typical simulations.

\begin{figure}[tb]
\centerline{\psfig{file=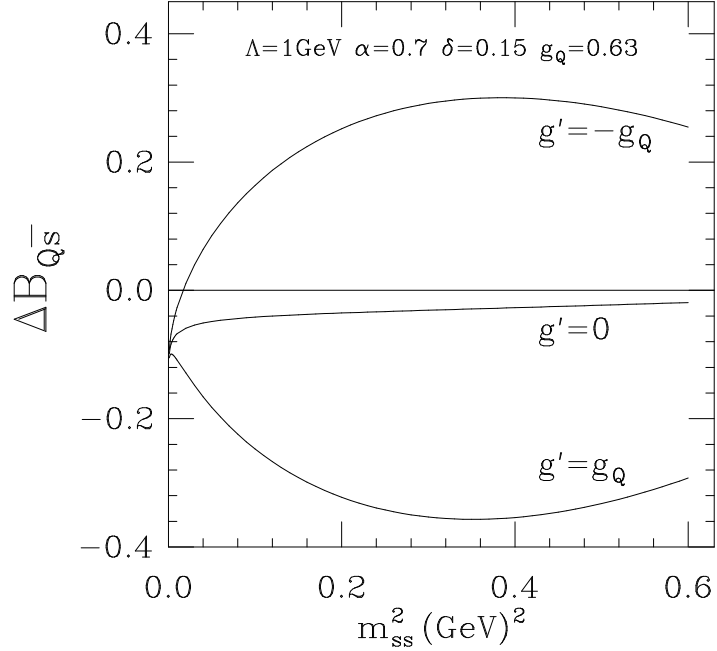,height=4.5truein}}
\vspace{-0.5truein}
\caption{Quenched chiral logarithmic contribution to $B_{Q\overline{s}}$.
Parameters as in Fig. 3.}
\label{fig:DBB}
\end{figure}

An example of the results for $(\Delta B_{Q\overline{s}})^{{\rm QQCD}}$,
is shown in Fig. \ref{fig:DBB}. There is very little dependence on
$\alpha$, so we show results only for $\alpha=0.7$.
Similar comments apply as for $\Delta f_{Q\overline{s}}$, except that
\begin{itemize}
\item
The sign of the term proportional to $g\prime$ has changed;
\item
The divergence at small $m_{ss}^2$ is weaker.
This is because it is proportional to $1- 3 g_Q^2$, to be compared
to $1+3 g_Q^2$ for $\Delta f_{Q\overline{s}}$.
The sign of the divergence depends on $g_Q^2$.
\end{itemize}

\subsection{Quantitative estimates of quenching errors}

These results suggest that quenched artifacts give
important contributions to $f_B$ and $B_B$.
They do not, however, allow a quantitative estimate of quenching errors.
For this we need to know something about the non-logarithmic
contributions. For example, to estimate $f_B^{\rm QQCD}/f_B^{\rm QCD}$ we
clearly
need to know both $\kappa^{{\rm QQCD}}/\kappa^{{\rm QCD}}$, and the difference
between the analytic $O(m_q)$ and $O(m_0^2)$ corrections.
These can compensate for differences in the chiral logarithms.

The cleanest way of proceeding is to consider ratios
which cancel as many of the unknowns as possible \cite{Golterman3}.
Single ratios such as $f_{B_s}/f_{B}$ (with $f_B^{\rm QQCD}$ defined by
extrapolation as explained above) do not depend on
$\kappa^{{\rm QQCD}}/\kappa^{{\rm QCD}}$, but do depend on the difference of
$O(m_q)$ terms in QCD and QQCD.
To remove this dependence, we must form a double ratio,
which requires the use of three non-degenerate valence quarks.
Calling these $u$, $d$ and $s$, and assuming $m_u < m_d < m_s$, we construct
\begin{eqnarray}
\label{eq:qf}
Q_f &=& \left( f_{Q\overline{u}} \over f_{Q\overline{d}}
		\right)^{m_s-m_u \over 2(m_d - m_u)}
      \left( f_{Q\overline{s}} \over f_{Q\overline{d}}
		\right)^{m_s-m_u \over 2(m_s - m_d)}
 - 1 \\
&=& {m_s-m_u \over 2(m_s - m_d)}
    (\Delta f_{Q\overline{s}} - \Delta f_{Q\overline{d}})
-   {m_s-m_u \over 2(m_d - m_u)}
    (\Delta f_{Q\overline{d}} - \Delta f_{Q\overline{u}}) + O(m_q^2) \,.
\label{eq:qfexpand}
\end{eqnarray}
and the similar ratio $Q_B$ involving the $B$-parameters.
To the order that we work in chiral perturbation theory, the
quark masses appearing in $Q_f$ can be replaced by the corresponding
squared meson-masses, e.g. $m_s \to m_{ss}^2$.
Any dependence of the form
$\Delta f_{Q\overline{q}} \propto a + b (m_u+m_d+m_s)
+c m_q$ cancels in $Q_f$, up to corrections of $O(m_q^2)$.
In particular $Q_f$ is independent of the cut-off $\Lambda$.
Note also that $Q_f$ has a well defined limit as $m_d\to m_s$ or $m_d\to m_u$.
An interesting special case is if the masses are in the ratio
$m_u:m_d:m_s=1:2:3$, for which
$Q_f= f_{Q\overline{u}} f_{Q\overline{s}} / (f_{Q\overline{d}})^2$.

$Q_f$ and $Q_B$ are thus quantities which we can calculate with some
reliability in all three theories which we consider.
The differences give a direct measure of the (partial) quenching error.
The results for QQCD follow from Eqs. \ref{eq:Dfds} and \ref{eq:Dbbs}
\begin{eqnarray}
Q_f^{\rm QQCD} & = & {m_{ss}^2 \over (4 \pi f)^2}
\left[(1 + 3 g_Q^2){\alpha\over3}
- \left( {i f V_L'(0) \over \sqrt6} + 3 g\prime g_Q\right) \right] R_1
- \delta {(1 + 3 g_Q^2) \over 4}  R_2 \,,\\
Q_B^{\rm QQCD} & = & {m_{ss}^2 \over (4 \pi f)^2}
\left[(1 - 3 g_Q^2)({2\alpha\over3}-1)  \right. \nonumber \\
&&\mbox{}+ \left.
{i 2 f\over \sqrt{6}} [V_O'(0)+V_L'(0)] + 6 g\prime g_Q \right] R_1
- \delta {(1 - 3 g_Q^2) \over 2} R_2 \,,
\end{eqnarray}
where $R_{1,2}$ are functions of only the quark mass ratios
\begin{eqnarray}
R_1 &=& {m_s - m_u \over 2 m_s} \left[
{m_s \log(m_s/m_d) \over m_s - m_d} - {m_u \log(m_d/m_u) \over m_d - m_u}
\right] \,, \\
R_2 &=& {m_s - m_u \over 2} \left[
{\log(m_s/m_d) \over m_s - m_d} - {\log(m_d/m_u) \over m_d - m_u}
\right] \,.
\end{eqnarray}
For QCD, after some algebra, we find
\begin{eqnarray}
Q_f^{\rm QCD} & = & - {(1+3 g^2) \over (4 \pi f)^2} {m_s-m_u \over 4}
\left[ {m_{\pi^+}^2 \log(m_{K^+}^2/m_{\pi^+}^2) \over m_s-m_d}
     - {m_{K^0}^2 \log(m_{K^0}^2/m_{K^+}^2) \over m_d-m_u} \right] \\
Q_B^{\rm QCD} & = & 0 \,.
\end{eqnarray}
The vanishing of $Q_B^{\rm QCD}$ seems to be accidental---we expect the
contributions of $O(m_q^2)$ to be non-vanishing.
Finally, for PQQCD, we have
\begin{eqnarray}
Q_f^{\rm PQQCD} & = & - {(1+3 g_{PQ}^2) \over (4 \pi f)^2}
\left[ {N_f \over 2} m_{ss}^2 R_3 + {1\over 2 N_f} m_{ff}^2 R_2
- {1\over N_f} m_{ss}^2 R_1 \right] \\
Q_B^{\rm PQQCD} & = &  - {(1-3 g_{PQ}^2) \over (4 \pi f)^2}
\left[ {N_f -2 \over N_f} m_{ss}^2 R_1 + {1\over N_f} m_{ff}^2 R_2 \right] \,,
\end{eqnarray}
where the additional ratio is
\begin{equation}
R_3 = {m_s - m_u \over 4 m_s} \left[
{m_s+m_f\over m_s - m_d} \log\left({m_s+m_f \over m_d+m_f}\right)
- {m_u+m_f \over m_d - m_u} \log\left({m_d+m_f \over m_u+m_f}\right)
\right] \,. \\
\end{equation}

\begin{table}[tb]
\begin{center}
\caption{\label{tab:Q}
	Results for ratios $Q$ in QCD and QQCD, in terms of
	$x=m_{ss}^2/0.5 {\rm GeV}^2$ and $y=\delta/0.15$,
	with $g\prime=V_L'(0)=V_O'(0)=0$.
        }
\begin{tabular}{lcccc}
$m_u:m_d:m_s$
   &$g^2=g_Q^2$	& $Q_f^{\rm QCD}$	& $Q_f^{\rm QQCD}$
				& $Q_B^{\rm QQCD}$ \\
\hline
$1:2:3$	& $0.4$	& $0.009\,x$	& $0.024\,\alpha\,x + 0.024\,y$
				& $0.004\,x\,(1.5\!-\!\alpha)-0.004\, y$ \\
$1:2:10$& $0.4$	& $0.026\,x$	& $0.082\,\alpha\,x + 0.182\,y$
				& $0.015\,x\,(1.5\!-\!\alpha) - 0.032\, y$ \\
$1:2:10$& $0.16$& $0.017\,x$	& $0.055\,\alpha\,x + 0.123\,y$
				&$-0.039\,x\,(1.5\!-\!\alpha) +0.086\,y$ \\
\end{tabular}
\end{center}
\end{table}

To make numerical comparisons we choose two sets of quark masses:
$m_u:m_d:m_s = 1:2:3$, ratios that are close to those
available from a typical simulation;
and $m_u:m_d:m_s = 1:2:10$, ratios now attainable in QQCD
for $m_s \approx m_s^{\rm phys}$\cite{Kim},
and which approach those of the real world.
The results are shown in Table \ref{tab:Q}.
Typical simulations have $x\approx1$, and we expect $y$ to lie
in the range $1/3-1$.
For the smaller mass ratios, we see that there is a difference between
$Q_f$ in QCD and QQCD, but the quenching error is likely not larger
than a few percent.
This is probably comparable to the statistical
errors in the best present simulations.
The corresponding error for $Q_B$ is much smaller, and unlikely to be
measurable.
Other values of the couplings lead to similar conclusions.

For the larger mass ratios, there is a substantial quenching error,
possibly as large as $20\%$ in $Q_f$. This is a reflection of the
divergence at small $m_{ss}$ in QQCD.
The quenching error in $Q_B$ is now larger.
but, as shown by the last column, is quite sensitive to $g_Q$.

We view these estimates as lower bounds on the quenching errors in
decay constants and $B$-parameters themselves, since the latter errors might
largely cancel in the ratios defining $Q_{f,B}$.
The estimates imply that one
should not simulate QQCD with quark masses that are too small.
They also suggest that quenching errors in $B$-parameters are
smaller than those in decay constants.

\begin{table}[tb]
\begin{center}
\caption{\label{tab:pqqcd}
	Results for ratios $Q$ in PQQCD,
        with $N_f=2$ and $g_{PQ}=0.63$.
        }
\begin{tabular}{lcccc}
	& \multicolumn{2}{c}{$Q_f^{\rm PQQCD}(m_{ss}^2=0.5{\rm GeV}^2)$}
	&\multicolumn{2}{c}{$Q_B^{\rm PQQCD}(m_{ss}^2=0.5{\rm GeV}^2)$} \\
$m_u:m_d:m_s$ 	& $m_{ff}^2=0.15 {\rm GeV}^2$
				& $m_{ff}^2=0.3 {\rm GeV}^2$
				& $m_{ff}^2=0.15 {\rm GeV}^2$
					& $m_{ff}^2=0.3 {\rm GeV}^2$ \\
\hline
$1:2:3$		& $0.021$	& $0.035$ & $-0.0015$	& $-0.003$\\
$1:2:10$	& $0.129$	& $0.216$ & $-0.012$	& $-0.024$\\
\end{tabular}
\end{center}
\end{table}

Similar conclusions hold also for ``partial quenching'' errors.
The results for $Q_f^{\rm PQQCD}$
are not linear functions of $m_{ss}^2$ and thus
more difficult to display. The values change little in the range
$m_{ss}^2=0.25-0.5 ({\rm GeV})^2$, however, so we quote the value
for $m_{ss}^2=0.5 ({\rm GeV})^2$.
There are no such problems for $Q_B^{\rm PQQCD}$,
since, for $N_f=2$, it is independent of $m_{ss}$ for fixed quark-mass ratios.
Our results are given in Table \ref{tab:pqqcd}.
They are much closer to the QQCD results than those from QCD.
For these quantities, partial quenching appears to be of little help.
The only caveat is that, as described in Sec. \ref{subsec:pq},
we have neglected terms suppressed by $(3/N_f)(m_{ff}^2/m_0^2)$, which
may not be small for the larger values of $m_{ff}^2$.

It would be interesting to test the results for $Q$ in quenched simulations.
One way of doing this would be to plot $Q_f/R_2$ versus $m_{ss}^2 R_1 / R_2$.
The results for all mass ratios should lie on a universal straight line,
the intercept of which is proportional to $\delta$.
An important issue facing such a test is the size of higher order chiral
corrections. To address this, we assume the correction effectively
adds to $\Delta f$ a contribution of size $(m_{ss}/ 1 {\rm GeV})^4$.
For $m_{ss}^2 = 0.5 {\rm GeV}^2$, this is a 25\% correction to
the decay constant itself.
The contribution to $Q_f$ turns out to be independent of $m_d$
\begin{equation}
\delta Q_f = {\textstyle{1\over2}}
 \left( {m_{ss}^2 -m_{uu}^2 \over 1 {\rm GeV}^2}
 \right)^2 \,.
\end{equation}
The importance of this correction depends on the quark masses.
Taking $m_{ss}^2 = 0.5 {\rm GeV}^2$, one has
$\delta Q_f=0.06$ for $m_u:m_d:m_s=1:2:3$,
which is larger than the supposedly leading terms.
On the other hand, $\delta Q_f=0.10$ for $m_u:m_d:m_s=1:2:10$,
which is smaller than the leading terms.
It seems likely, though, that one would need to work at
somewhat smaller quark masses to suppress such $O(m_q^2)$ terms.

\subsection{Qualitative estimates of quenching errors}

We turn now to the quenching errors in the individual quantities
$f_B$, $f_{B_{s}}$, $B_B$ and $B_{B_{s}}$.
The notation we use is exemplified by
\begin{equation}
\Delta(\Delta f_B) = f_B^{\rm QQCD}/f_B^{\rm QCD} - 1 \,.
\end{equation}
In the previous subsection, we used QCD to refer to all theories with
three light quarks, irrespective of their masses.
In this subsection, however, QCD means the theory with physical quark masses.
To make this distinction clear, we use $f_{B_{s}}$
instead of $f_{Q\overline{s}}$
(and $B_{B_{s}}$ in place of $B_{Q\overline{s}}$),
although the two quantities are the same.
We define the (partially) quenched $f_B$ and $B_B$
by linear extrapolation in the light quark mass, as explained above.

We estimate quenching errors by assuming that the dominant difference between
quantities in QQCD and QCD comes from the chiral logarithms.
In other words, we assume, for some reasonable choice of $\Lambda$, that
\begin{equation}
\Delta(\Delta f_B) \approx
(\Delta f_B)^{\rm QQCD}(\Lambda) - (\Delta f_B)^{\rm QCD}(\Lambda) \,,
\end{equation}
and similarly for the other quantities.
This assumption is likely to be wrong for individual quantities,
because differences in chiral logarithms can be compensated
by differences in analytic contributions.
(By analytic contributions we mean not only the $O(m_q)$ terms
proportional to $c_{1,2}$, but also the lowest order terms,
e.g. $\kappa^{{\rm QQCD}}$.)
One might expect there to be some compensation because
the two theories are matched by equating, say, $f_\pi$,
and so the difference in chiral logarithms for this quantity are, perforce,
compensated.
But it seems unlikely to us that this compensation will occur uniformly;
for one thing, the chiral logarithms due to $\eta'$ loops are absent
for some quantities (e.g. $f_\pi$), while present for others.
Thus we view our estimate of the quenching error as a rough upper
bound (because there may be some compensation), which is likely to work
better on the average over several quantities.


\begin{table}[tb]
\begin{center}
\caption{\label{tab:delta}
	Numerical results for chiral logarithms $\Delta^{\rm QQCD}$ and
	$\Delta^{\rm QCD}$, the latter being in parentheses. Our estimate of
	the quenching error is the difference of these two expressions.
	We assume $V_L'(0)=V_O'(0)=0$.
        }
\begin{tabular}{lcc}
Qty.              & $\Lambda=1\,$GeV & $\Lambda=0.77\,$GeV \\
\hline
$f_{B_{s}}$       & $-0.095\alpha + 0.38\delta + 0.25g\prime$ \, ($0.38$) &
          $-0.023\alpha + 0.093\delta+ 0.060g\prime$ \, ($0.23$) \\
$f_B$           & $-0.095\alpha + 1.14\delta + 0.25g\prime$ \, ($0.21$)  &
          $-0.095\alpha + 0.85\delta + 0.25g\prime$ \, ($0.14$) \\
\makebox[0pt][l]{${f_{B_{s}} / f_B}$}
	& $-0.0\alpha - 0.76\delta+ 0.0g\prime$ \, ($0.17$)     &
          $-0.072\alpha - 0.76\delta - 0.19g\prime$ \, ($0.09$) \\
$B_{B_{s}}$   & $-0.025+0.017\alpha-0.066\delta-0.49g\prime$ \,($-0.017$) &
          $-0.006+0.004\alpha-0.016\delta-0.12g\prime$ \,($-0.010$) \\
$B_B$       & $-0.025+0.017\alpha-0.20\delta-0.49g\prime$ \, ($-0.007$) &
          $-0.025+0.017\alpha-0.15\delta-0.49g\prime$ \,($-0.005$) \\
\makebox[0pt][l]{${B_{B_{s}} / B_B}$}
	& $-0.0+0.0\alpha+0.13\delta-0.0g\prime$ \,($-0.010$)   &
          $0.019-0.013\alpha+0.13\delta+0.37g\prime$ \,($-0.005$) \\
\end{tabular}
\end{center}
\end{table}

\begin{figure}[tb]
\centerline{\psfig{file=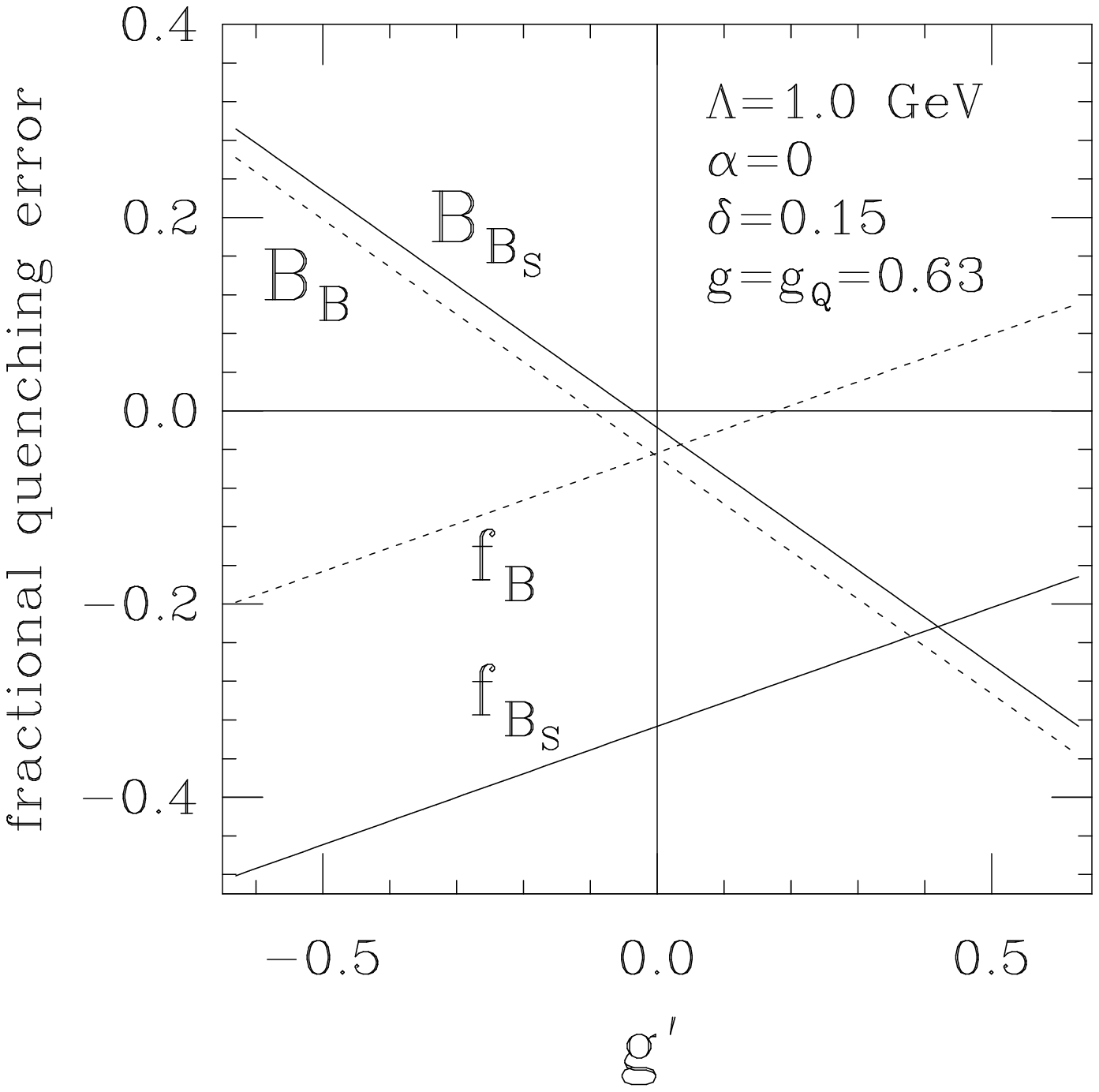,height=5.truein}}
\vspace{-0.5truein}
\caption{Results for $\Delta(\Delta X)$ for $\Lambda=1\,$GeV, $\alpha=0$,
$\delta=0.15$ and $g=g_Q=0.63$.}
\label{fig:qerr4}
\end{figure}

\begin{figure}[tb]
\centerline{\psfig{file=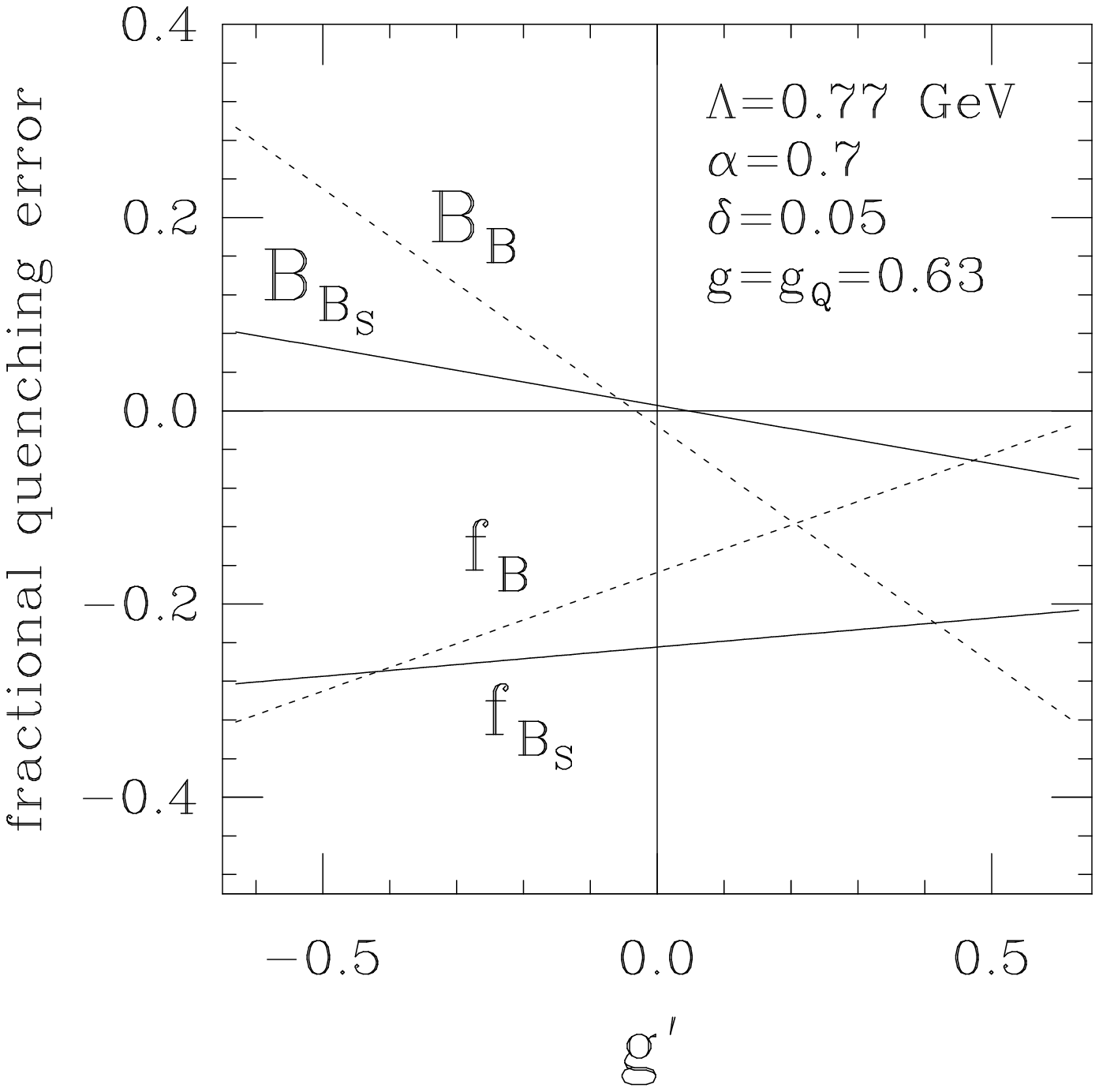,height=5.truein}}
\vspace{-0.5truein}
\caption{Results for $\Delta(\Delta X)$ for $\Lambda=0.77\,$GeV, $\alpha=0.7$,
$\delta=0.05$ and $g=g_Q=0.63$.}
\label{fig:qerr3}
\end{figure}

We have examined the resulting estimates for our reasonable ranges of
parameters,
and for the choices $\Lambda=m_\rho$ and $\Lambda=1\,$GeV.
To illustrate their size we give in Table \ref{tab:delta} the numerical
expressions for the chiral logarithms
in terms of $\alpha$, $\delta$ and $g\prime$, fixing $g=g_Q=0.63$.
In Figs. \ref{fig:qerr4} and \ref{fig:qerr3} we plot the quenching
error versus $g\prime$ for two sets of parameters chosen to
indicate the range of variation in the estimates.

We draw the following conclusions.
\begin{itemize}
\item
There are no values of the parameters for which the quenching errors
are small simultaneously for all four quantities,
where small means $|\Delta(\Delta X)|<0.1$.
Typically, two or more of the estimates are larger than 0.2 in magnitude,
some being as large as 0.4.
Viewing our estimates as upper bounds, we conclude that
it is possible that there are substantial quenching errors in one or
more of the quantities.
\item
It is interesting to consider the ratios $f_{B_{s}}/f_B$ and $B_{B_{s}}/B_B$.
We expect our estimate of the quenching errors to be more reliable here,
since there is no dependence on the leading chiral parameters, e.g.
$\kappa^{\rm QQCD}/\kappa^{\rm QCD}$.
One might also hope that quenching errors are smaller.
Indeed, there are cancellations in the chiral logarithms,
as one can see from Table \ref{tab:delta}.
(The exact cancelation of the coefficients of $\alpha$ and $g\prime$
for these ratios when $\Lambda=1\,$GeV is {\em not} a general
feature---it is a chance result, valid only for
the particular masses we use to extrapolate to the $B$-meson.)
Nevertheless, as is clear from Figs. \ref{fig:qerr4} and \ref{fig:qerr3},
the error in $f_{B_{s}}/f_B$ (given by the difference of the curves
for these two quantities) can still be substantial.
\item
There is a considerable dependence of the estimates on the unknown
coupling $g\prime$. If we trust the large $N_c$ expansion, and assume
$|g\prime/g_Q|<0.3$, then we can draw further conclusions. In particular:
\begin{itemize}
\item
For small $g\prime$, the estimated errors in $B_B$ and $B_{B_{s}}$ are small,
less than roughly 0.1 in magnitude. This is true for all values of $g_Q^2$
in our reasonable range.
The error in the ratio $B_{B_{s}}/B_B$ is smaller still, though of either sign.
\item
The error in the ratio $f_{B_{s}}/f_B$ is negative, and lies in
range $-0.3$ to $-0.1$. As discussed above, this is because the quenched
chiral logarithm decreases with $m_{ss}$, due to the term
proportional to $\delta$.
Taken literally, this would imply that the ratio
in QCD is 10-30\% larger than in QQCD.
More conservatively, it suggests that the deviation of the ratio from unity,
which in present QQCD simulations is $(f_{B_{s}}/f_B-1) \sim 0.1-0.2$,
has a quenching error of 100\% or more.
\end{itemize}
\end{itemize}

\begin{table}[tb]
\begin{center}
\caption{\label{tab:ddpq}
	Numerical results for $\Delta^{\rm PQQCD}-\Delta^{\rm QCD}$,
	with $g^2=g_{PQ}^2=0.4$.
        }
\begin{tabular}{lcccc}
Qty.           	& \multicolumn{2}{c}{$\Lambda=1\,$GeV}
		& \multicolumn{2}{c}{$\Lambda=0.77\,$GeV} \\
	& $m_{ff}^2=0.15 {\rm GeV}^2$ &$m_{ff}^2=0.3 {\rm GeV}^2$
	& $m_{ff}^2=0.15 {\rm GeV}^2$ &$m_{ff}^2=0.3 {\rm GeV}^2$ \\
\hline
$f_{B_{s}}$	&$-0.20$	&$-0.18$	&$-0.10$	&$-0.13$ \\
$f_B$  	&$-0.06$	&$ 0.05$	&$-0.04$	&$ 0.03$ \\
\makebox[0pt][l]{${f_{B_{s}}/ f_B}$}
	&$-0.14$	&$-0.24$	&$-0.06$	&$-0.16$ \\
$B_{B_{s}}$ &$ 0.014$	&$ 0.010$	&$ 0.009$	&$ 0.008$ \\
$B_B$ 	&$-0.004$	&$-0.015$	&$-0.004$	&$-0.012$ \\
\makebox[0pt][l]{${B_{B_{s}} / B_B}$}
	&$ 0.018$	&$ 0.025$	&$ 0.012$	&$ 0.020$
\end{tabular}
\end{center}
\end{table}

We show in Table \ref{tab:ddpq} our estimates of errors due to
partial quenching.
The results are for $g^2=g_{PQ}^2=0.4$, but it is simple to
scale to other values of $g$ because the overall factors are the
same for QCD and PQQCD.
The results are fairly similar to those in Fig. \ref{fig:qerr3} for
$g\prime\approx0$. Most notably, the partial quenching error
in $f_{B_{s}}/f_B$ is negative and significant, just as for QQCD.
The errors in $B$-parameters are very small.

\section{Conclusions}

We have presented estimates for errors in lattice results for
decay constants and $B$-parameters of heavy-light mesons
due to quenching and partial quenching.
Our estimates are reliable for the ratios $Q_{f,B}$, aside from
higher order chiral corrections, or, in the case of PQQCD,
corrections suppressed by $m_{ff}^2/m_0^2$.
Our estimates for the individual quantities, however,
should be treated as approximate upper bounds.
They suggest that, in order to extract $f_B$ and $B_B$,
it is better to extrapolate from moderate quark masses than to attempt
to work directly with very light valence quarks.
As for the quenching errors themselves,
those for $B$-parameters are likely
to be smaller than those in decay constants,
with the latter possibly as small as a few percent,
but more likely larger.
Our most striking finding is that quenching
reduces the ratio $f_{B_{s}}/f_B$, throwing doubt on the common assumption
that this ratio is a fairly reliable prediction of QQCD.

Finally, we have found similar results for partially quenched QCD
as for the fully quenched theory. This extends the observation of
Booth \cite{Booth1,Booth2}, who found that there were significant
differences between results for the unquenched theory with two flavors
and QCD.

\section*{Acknowledgements}
We thank Claude Bernard and Maarten Golterman for helpful
discussions and for comments on the manuscript.

\end{document}